\overfullrule=0 pt
\def\jcd{Christensen-Dalsgaard}

\input epsf
\newdimen\tabledimen  \tabledimen=\hsize
\def\table#1#2{\tabledimen=\hsize \advance\tabledimen by -#1\relax
\divide\tabledimen by 2\relax\vskip 1pt
\moveright\tabledimen\vbox{\tabskip=1em plus 4em minus 0.9em
\halign to #1{#2}} }
%
\def\tabmidrule{\noalign{\smallskip\hrule\smallskip}}             
\def\eqa#1{\global\advance\eqnum by 1\relax
\xdef#1{{\noexpand{\rm}(\number\eqnum)}}
{\rm(\the\eqnum)}}
\input aa.cmm
\voffset 0.5 cm
\MAINTITLE{Estimate of solar radius from f-mode frequencies}
\AUTHOR{H. M. Antia}
\INSTITUTE{
Tata Institute of Fundamental Research,
Homi Bhabha Road, Mumbai 400 005, India}
\OFFPRINTS{H. M. Antia}
\DATE{Received \ \ , accepted \ \ }

\ABSTRACT{Frequency and rotational splittings of the solar f-modes are
estimated from the GONG data.
Contrary to earlier observations the frequencies
of f-modes are found to be close to the theoretically computed
values for a standard solar model.
The f-mode being essentially a surface mode is
a valuable diagnostic probe of the properties of the
solar surface, and also provides an independent measure of solar radius.
The estimated solar radius is found to be about 0.03\% less than what is
traditionally used in construction of standard solar models.
If this decrease in solar radius is confirmed then the current solar models
as well as inversion results will need to be revised.
The rotational splittings of the f-modes yield an independent measure
of the rotation rate near the solar surface, which is compared with
other measurements.}

\KEYWORDS{ Sun: oscillations  -- Sun: rotation}
\THESAURUS{09(06.15.1; 06.18.2)}
\maketitle

\titlea{Introduction}

The frequencies and splittings of the solar $p$-modes have been
extensively and profitably used in helioseismic analysis to infer
conditions in
the solar interior. The f-mode which is essentially a surface
mode has also attracted attention because of the reported difference
in the frequency between the observed value and that computed
for a solar model (Libbrecht, Woodard \& Kaufman 1990; Bachmann et al.~1995).
Since frequencies of the f-mode are essentially independent of
the stratification in the solar interior, they can provide a
diagnostic of flows and magnetic fields etc. present in the
near surface regions
(Murawaski \& Roberts 1993; Rosenthal \& Gough 1994;
Ghosh, Chitre \& Antia 1995; Rosenthal \& \jcd\ 1996). 
These frequencies can also provide an accurate measure of solar radius.

The amplitudes of f-modes are
very low, and consequently, the frequencies have so far been measured
only at high
degree where there is sufficient power. These frequencies probably
suffer from systematic errors (Antia 1996), presumably because of
ridge fitting techniques adopted in data reduction.
It may also be noted that for the f-mode, horizontal and vertical
components of velocity are comparable in magnitude and the usual
assumption in spatial filtering about velocity being predominantly vertical
is untenable. Systematic errors introduced
because of this assumption have perhaps not been estimated.
 However, with the good
quality data now available from the GONG network (Hill et al.~1996)
and the MDI project (Kosovichev and Schou 1997), it is possible
to detect the fundamental mode down to approximately $\ell=100$,
where $\ell$  is the degree of the mode.
The advantage with the
GONG and MDI data is that they provide information about individual modes
and hence ridge fitting is not involved. Thus we can expect the systematic
errors in the estimated frequencies to be much less.

Rest of the paper is organized as follows: In section 2, we describe the
results obtained for the frequencies of the f-mode from the GONG data
including the estimate of solar radius,
while section 3 describes the results for splitting coefficients for the
f-modes and the surface rotation rate as inferred from them. Finally,
section 4 summarizes the main conclusions from this study.

\titlea{the f-mode frequencies}

To determine the frequencies of f-modes from GONG power spectra
it is best to use the rotationally corrected $m$-averaged power spectra;
because of addition of spectra for all value of $m$ the signal to
noise ratio is improved and it is possible to identify the f-mode
peaks without much difficulty for $\ell>100$. We use the
GONG month 4 power spectra because these were available for $\ell=0$--250.
From these $m$-averaged spectra the frequencies have been found using
the standard peak-finding technique in the GONG pipeline
(Anderson, Duvall \& Jefferies 1990).

\begfigwid 10.0 cm
\moveleft -1.0 cm\vbox to 0 pt{\vskip -14.9 cm
\epsfysize=20.90 cm\epsfbox[1 1 615 792]{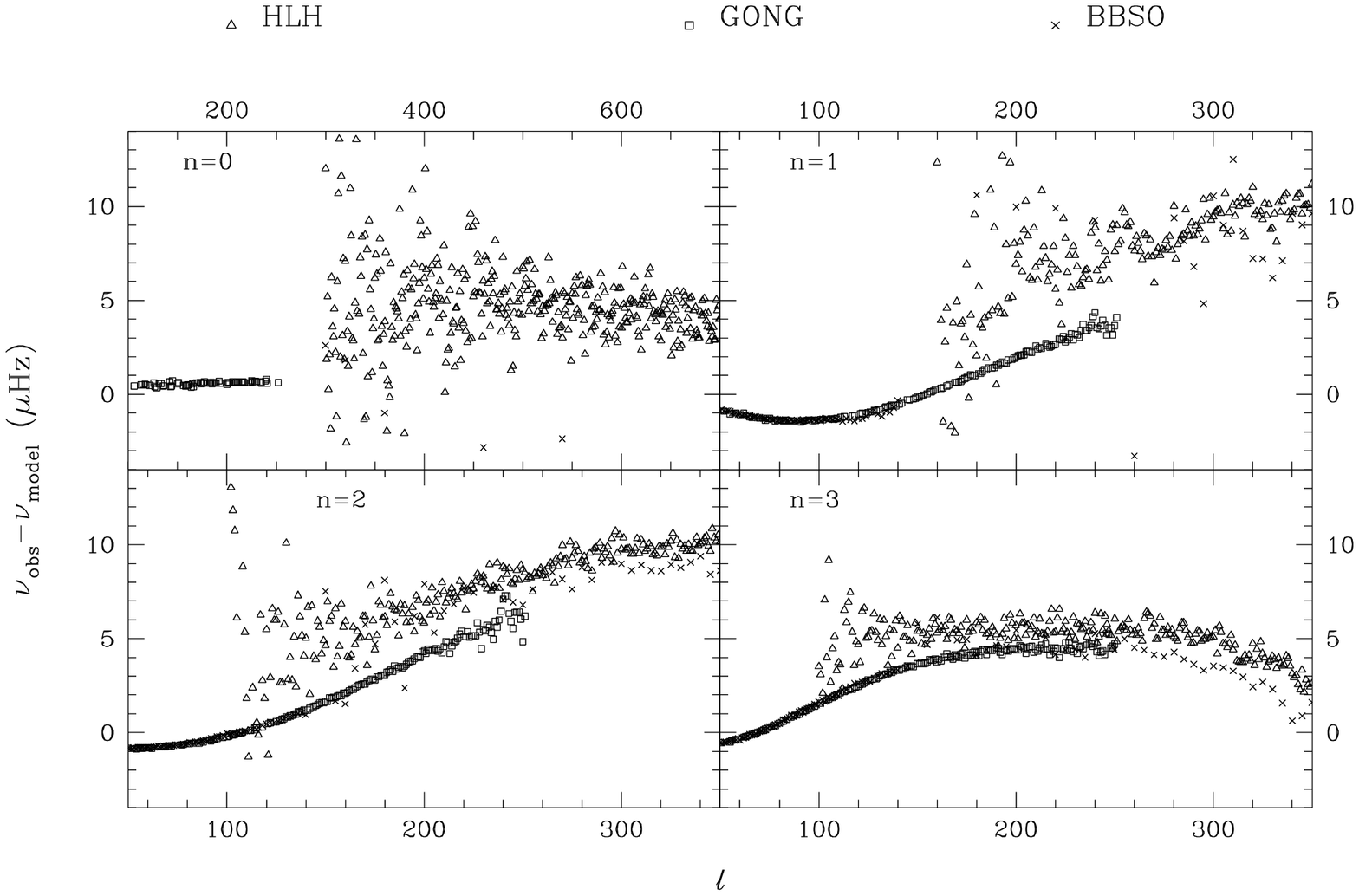}\vskip -5.0 cm}
\figure{1}{Difference between various observed frequencies and those
of a standard solar model for $n=0$--3 as a function of degree $\ell$.}
\endfig

The results for modes with radial harmonic
number $n=0$--3 are shown in figure 1. Although in this work we are
mainly interested in the f-mode ($n=0$), results for other $n$'s are also
included to show the gradual variation in results with $n$. These
figure shows the difference between the observed frequencies and those
of a standard solar model. Further, in order to estimate systematic
errors between different observations we have included observed frequencies
from the BBSO data (Libbrecht, Woodard \& Kaufman 1990) as well as
the HLH data (Bachmann et al.~1995). The BBSO data falls in two categories
one for $\ell\le140$ where the frequencies have been determined by fitting
individual peaks, similar to what is done for the GONG data,
and the second set for higher $\ell$ where the frequencies
have been computed by a ridge fitting technique. The HLH frequencies
have all been computed using ridge fitting. It is clear from the
figure that for $n=0$ and 1, there are significant differences between
the frequencies computed by fitting individual modes and those from
ridge fitting. Further, the frequencies from GONG and BBSO ($\ell\le 140$)
data are very close to those of the standard solar model, while those
from ridge fitting techniques are systematically different. The
difference between different data sets reduces as one goes to higher
values of $n$. The thickness of the ridges in these figures should give
an estimate of statistical errors in observed frequencies and it is clear that
the errors in GONG data are much less than those in HLH data.

The f-mode frequencies from GONG data are close
to those computed with a standard solar model and various mechanisms invoked to
explain the reported differences between the observed and computed
frequencies
may not be necessary. Of course, we still do not have
reliable results at high degree and only better data from MDI or a
reanalysis of HLH data would be able to resolve the question whether
there is indeed any significant difference between the observed and
computed frequencies. From Figure~1, it appears that there is still a
small difference of the order of $0.5\;\mu$Hz between the observed and
computed frequencies.

From the behavior of systematic error with
$n$ it appears that the neglect of horizontal component of velocity in spatial
filtering is a possible cause for the systematic differences.
 From the computed eigenfunction we
can obtain the ratio of the horizontal to vertical component of velocity at the
photosphere. This ratio is unity for $n=0$, between 0.4--0.5
for $n=1$, between 0.24--0.32 for $n=2$ and between 0.15--0.25 for $n=3$.
The influence of horizontal component of velocity on ridge fitting
techniques needs to be further studied.
It may be noted that for the $p$-modes the ratio of horizontal to vertical
velocity will be a function of frequency and as such spatial filtering
could introduce asymmetry in the peaks thus causing a shift in the
frequency when symmetric profiles are fitted (Kosovichev et al.~1997).

\begfig 8.5 cm
\moveleft 1.8 cm\vbox to 0 pt{\vskip -11.8 cm
\epsfysize=15.70 cm\epsfbox[1 1 615 792]{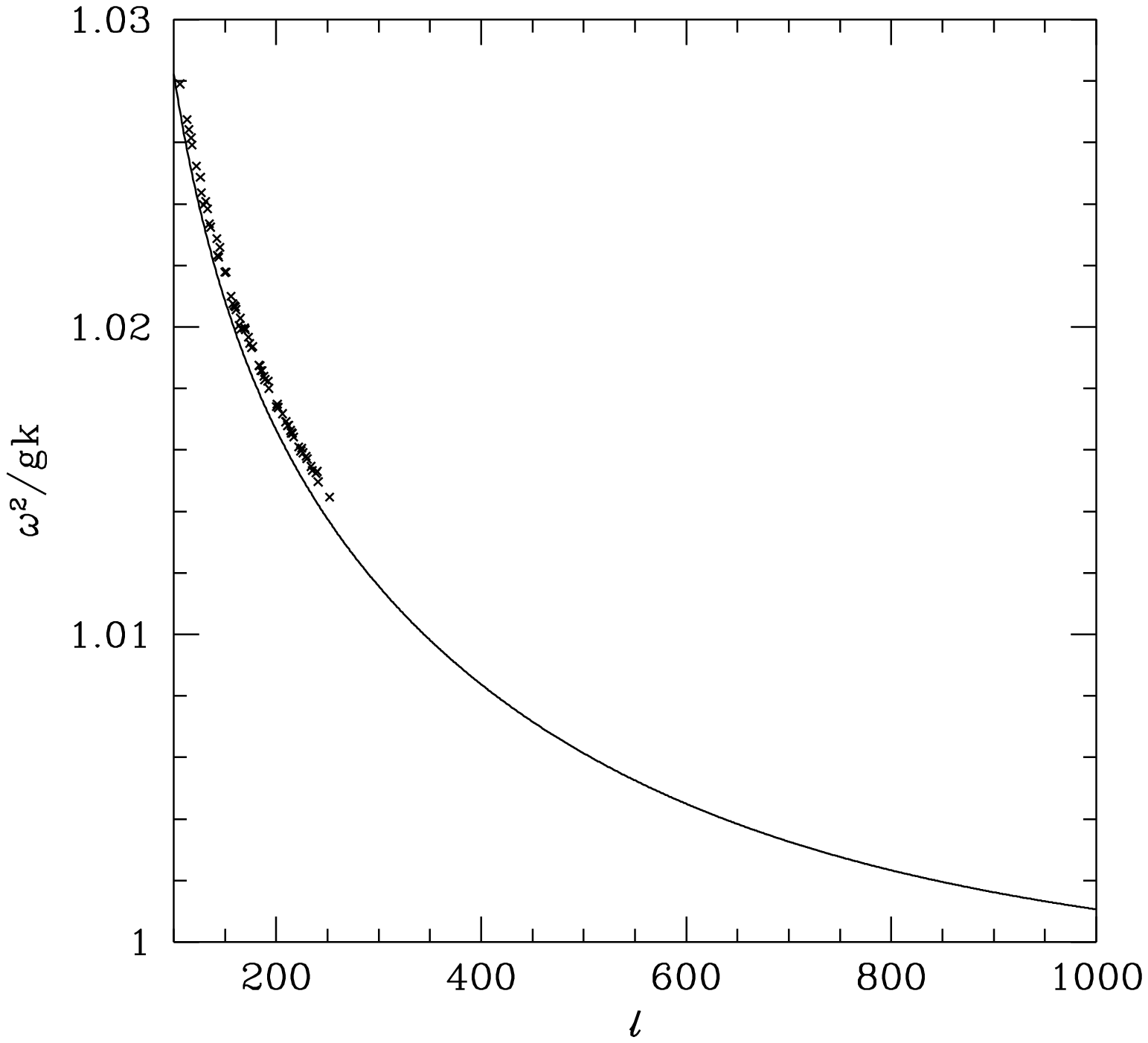}\vskip -3.9 cm}
\figure{2}{ $\omega^2/gk$ for the
f-mode in a solar model (solid line) are compared
with observed values represented by crosses.}
\endfig

In the ridge fitting techniques where the $\ell\pm1$ leaks in the power
spectra are not resolved the systematic errors are found to be of
the order of frequency separation $(\nu_{l+1,n,m}-\nu_{l,n,m})$. Thus in the
GONG data where the $m\pm2$ leaks are not resolved we may expect systematic
errors of the order of separation between these peaks, which is consistent
with the actual difference of the order of $0.5\mu$Hz seen between the
observed and computed frequencies
for a solar model. Thus it is possible that this
difference is again due to systematic errors in observed frequencies.
However, in that case the frequency difference would be independent of
$\ell$, but the actual difference is more or less proportional to
the frequency. It thus appears
unlikely that most of the difference could be accounted for by systematic
errors in measured frequencies. Hence in the following section we neglect
the possibility of unknown systematic errors and investigate the
consequences.

\begfig 8.5 cm
\moveleft 1.5 cm\vbox to 0 pt{\vskip -11.8 cm
\epsfysize=15.40 cm\epsfbox[1 1 615 792]{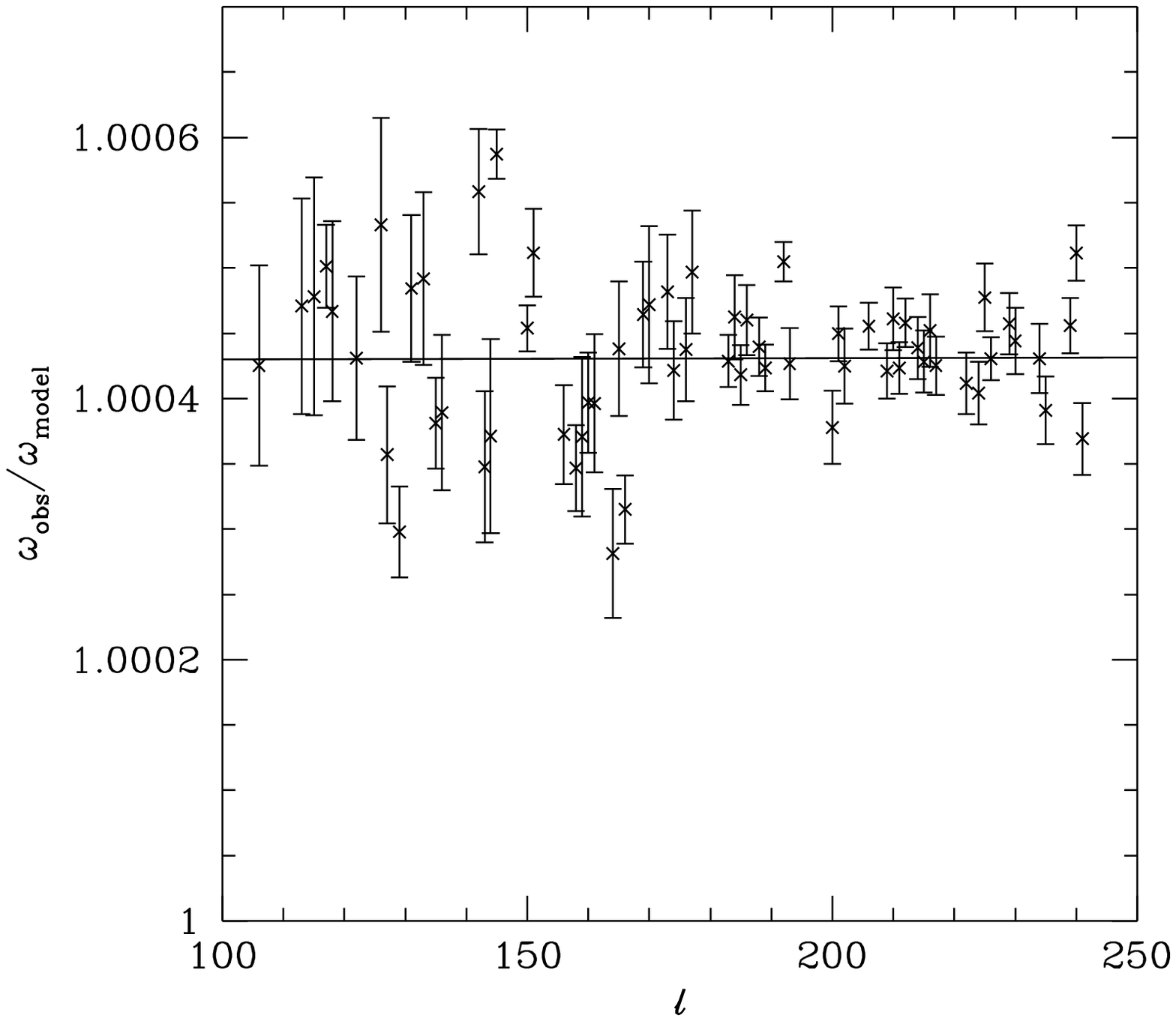}\vskip -3.6 cm}
\figure{3}{The ratio of observed and model frequencies for f-modes. The
horizontal line defines the average over all modes.}
\endfig

\titleb{Estimate of solar radius}

The frequencies of f-modes are asymptotically expected to satisfy the
simple dispersion relation, $\omega^2=gk$, where $g$ is the acceleration
due to gravity at the surface and $k=\sqrt{\ell(\ell+1)}/r$ is the
horizontal wave number. Figure~2 shows the quantity $\omega^2/gk$
for a solar model and for the corresponding GONG frequencies.
It is clear that
although both follow the same trend, there are some systematic
differences between the two.
The systematic trend away from unity at lower
degree is due to the fact that the peak in kinetic energy
density associated with f-mode shifts inwards with decreasing degree
and thus these modes are effectively localized somewhat below the
solar surface, where $gk$ would be larger. This arises because
although the velocity falls off exponentially with increasing depth,
the density increases very rapidly just below the solar surface.
As a result, the kinetic energy density increases until the density
scale height becomes comparable to velocity scale height ($1/k$).
Figure~3  shows the ratio
$(\omega_{\rm obs}/\omega_{\rm model})$ and it is clear that this
ratio is more or less constant within the expected errors. Moreover
it is significantly different from unity, with the average ratio being
$1.000437\pm0.000005$. The simplest explanation for this difference
in frequencies would be an error in the assumed radius of the
solar model. In order to explain the observed discrepancy the solar
radius will need to be decreased by about 0.029\% or about 203 km, which
is perhaps somewhat larger than the quoted uncertainty of 70 km in the radius.

However, there is a significant variation in measured value of the solar
radius, both with time and with different observational techniques
(Laclare et al.~1996).
Thus a reduction of 203 km in present solar radius cannot be ruled out.
Of course, some of the difference could arise from the assumed definition
of solar radius. For the present study, the solar model was constructed
with a radius of $6.9599\times10^5$ km, and the radius was defined as the
radial distance at which the temperature equals the effective temperature.
This point would be about 50 km above the level where the optical depth
equals unity. However, the definition of radius as used by observers
is quite different as they measure the distance to the 
inflection point of the limb intensity profile, which probably occurs
at a much lower optical depth. Wittmann (1974) has estimated this point
to be 340 km above the level where optical depth is unity.
Thus, the reduction by 203 km in solar radius
suggested by the f-mode frequencies appears to be roughly consistent with
the standard value of radius.

Of course, there could be other sources
to explain the difference between observed and model frequencies,
(Murawaski \& Roberts 1993; Rosenthal \& Gough 1994;
Ghosh, Chitre \& Antia 1995; Rosenthal \& \jcd\ 1996)
but these will again yield
a different behavior of differences with $\ell$. Since the observed
relative difference is essentially independent of $\ell$, even when
$\ell$ varies by more than a factor of two, it appears that a dominant
contribution to this difference is coming from the error in radius.
With the quality of data presently available it does not appear to be
possible to separate out the contributions from various possible sources
to the measured frequency differences. In any case before the f-mode
frequencies can be used to draw inference on any of these effects it
is essential to determine the solar radius correctly.

With better data on the f-mode becoming available, it may be possible
to estimate the value of solar radius more accurately
as also its possible variation with solar cycle.
Since the frequencies
of these modes can be determined to a relative accuracy of $10^{-5}$,
in principle, it would be possible to
determine the solar radius to much better accuracy.

It may be noted that most of the current standard solar models
(e.g., \jcd\ et al.~1996) use the standard value of solar radius with
the surface defined at a level where optical depth is between 1 and 1/3
and thus these models need to be revised. Similarly, most helioseismic
inversions assume a similar definition of solar radius and will also need
to be revised. In order to estimate the possible errors due to uncertainty
in radius we have tried helioseismic inversions for the sound speed
using the GONG months 4--10 data with different estimates of radius.
For this purpose we use a regularized least squares technique (Antia 1996)
with two different reference models M0 and M1, using identical physics and
identical composition profiles but with
radius 695990 and 695780 km respectively. In order to compare the
two results the relative difference with respect to the reference models
is converted to that with respect to model M1 and the results are shown
in Figure~4. This figure shows the relative difference taken at the
same fractional radius. It is clear that the difference caused due to
a change of radius by 210 km, is much more than the estimated errors in
helioseismic inversions over most of the solar interior. Clearly, we need
an accurate measure of the solar radius in order to infer the conditions
in solar interior accurately.

\begfig 8.5 cm
\moveleft 1.5 cm\vbox to 0 pt{\vskip -11.8 cm
\epsfysize=15.40 cm\epsfbox[1 1 615 792]{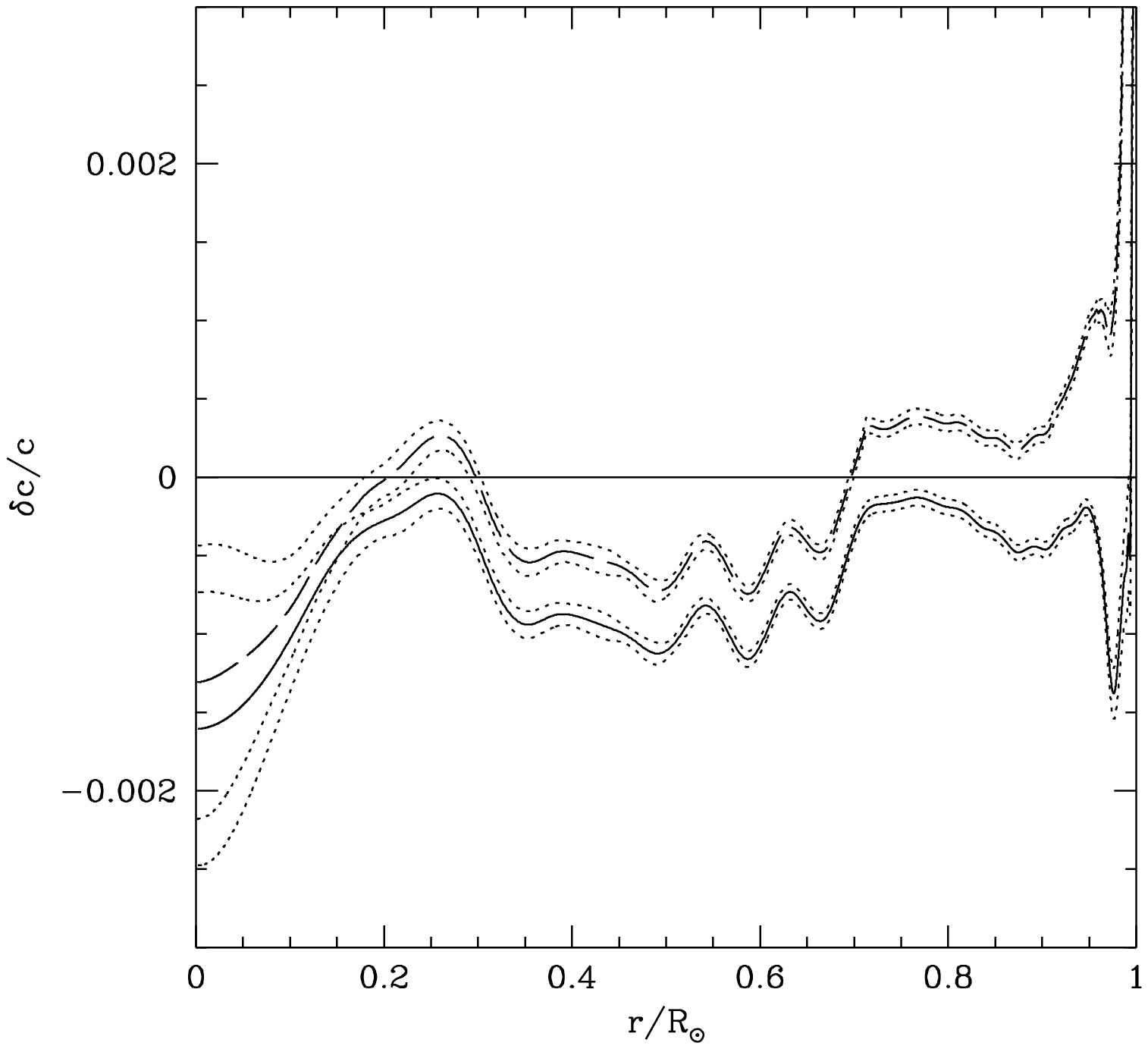}\vskip -3.6 cm}
\figure{4}{The relative difference in sound speed between the Sun and
model M1 as inferred by helioseismic inversions using two different estimates
for radius. The continuous and dashed lines show the result using 
$R_\odot=695780$ and 695990 km respectively. The dotted lines represent
the $1\sigma$ error limits on these inversions.}
\endfig

\titlea{The f-mode splitting coefficients}

Apart from the frequencies it is also possible to estimate the rotational
splitting coefficients for the f-modes.  For this purpose we
have used the results from GONG spectra averaged over several months, which
are available up to $\ell=150$ only. 
We attempt to calculate the mean frequency
and the first five splitting coefficients, using a least squares
fit to polynomials of Ritzwoller and Lavely~(1991),
$$\nu_{n\ell m}=\nu_{n\ell }+\sum_{i=1}^5 c_{i,n\ell }\gamma_{i,\ell }(m),
 \eqno(1)$$
where, $\gamma_{i,\ell}(m)$ are the polynomials defined by Ritzwoller \&
Lavely, $\nu_{n\ell}$ is the mean frequency and $c_{i,n\ell}$ are the
splitting coefficients.

Since all the
f-modes are restricted to a narrow region just below the surface, we
would expect the splitting coefficients to be roughly independent of $\ell$.
Thus it is possible to take 
mean of all these values and obtain more accurate splitting coefficients.
The average values for the first five coefficients are listed in
Table~1, which
shows the results obtained using different averaged spectra.
It can be seen that the three results are reasonably close to each other.

\begtabfull
\tabcap{1}{Mean Rotational Splitting coefficients for f-modes}
\table{\hsize}{&\hfil$#$\hfil\cr
\tabmidrule
&\hbox{Months 4--7}&\hbox{Months 4--8}&\hbox{Months 4--10}\cr
\tabmidrule
c_1&895.4\pm0.6   &895.5\pm0.5& 895.1\pm0.3\cr
c_2&-0.041\pm0.010&-0.007\pm0.010&0.003\pm0.004\cr
c_3&-18.4\pm0.4   &-18.5\pm0.4& -18.8\pm0.2\cr
c_4&-0.009\pm0.012&-0.003\pm0.013& 0.005\pm0.006\cr
c_5&-1.8\pm0.3    &-1.0\pm0.4& -1.7\pm0.2\cr
\tabmidrule}
\endtab

\begfig 8.6 cm
\moveleft 1.6 cm\vbox to 0 pt{\vskip -11.9 cm
\epsfysize=15.90 cm\epsfbox[1 1 615 792]{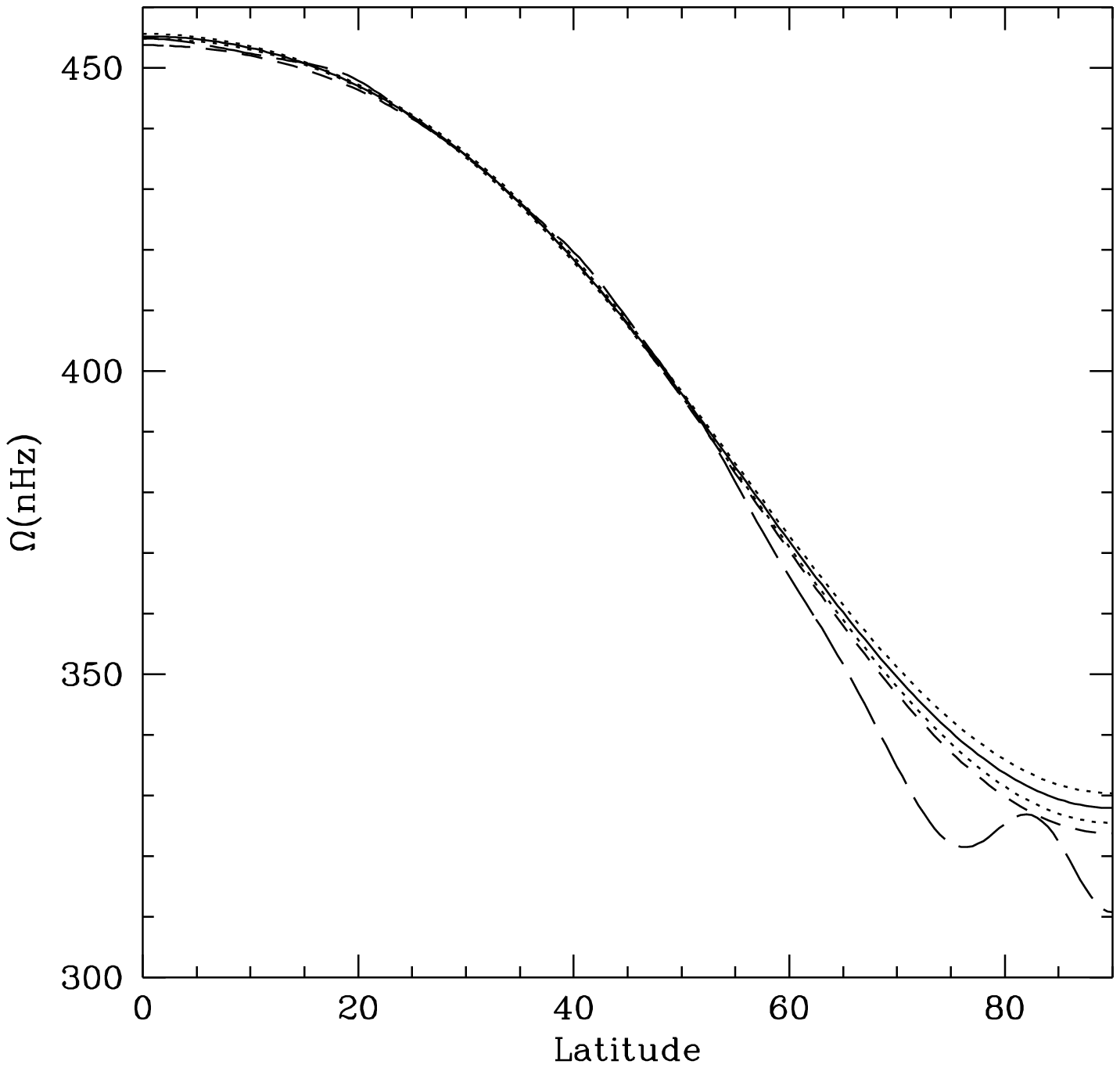}\vskip -4.0 cm}
\figure{5}{  The solar surface rotation rate
as inferred from the f-mode splittings (solid line) is compared with
other measurements. The dotted lines represent $1\sigma$ errors in the
rotation rate, while the short dashed line represents the surface rotation
rate as inferred by Doppler shifts (Snodgrass 1992) and the long dashed line
represents the surface rotation rate as inferred by inversion of GONG
months 4--10 data}
\endfig

Since the f-modes are confined to layers immediately below the solar surface,
we would expect that the splitting
coefficients would directly give the corresponding components of the
rotation rate.
The results using the coefficients as determined from the GONG months 4--10
data are shown in Figure~4, which also shows the solar surface
rotation rate as inferred from doppler measurements (Snodgrass 1992).
This is consistent with the results obtained from MDI data (Kosovichev
\& Schou 1997).  This figure also
shows the surface rotation rate as inferred by proper inversion of all
splitting coefficients from the GONG months 4-10 averaged spectra. It can
be seen that the inverted rotation rate is fairly close to that inferred
directly from the f-mode, and further some of the difference is due
to inclusion of higher coefficients in inversion (i.e., $c_7$ -- $c_{35}$).
From this figure it is clear that there is a reasonable agreement in surface
rotation rate inferred from various techniques, though the differences
are possibly larger than estimated errors.

\titlea{Conclusions}

The frequencies of the solar f-mode as determined from the GONG
power spectra for $100\le\ell\le250$ are reasonably close to those of a
standard solar model and it appears that a significant fraction of
the discrepancy noted in earlier observations is
due to systematic errors in estimating the frequencies from the observed
power spectra. Large horizontal component of velocity for the
f-mode could be a possible source of systematic errors in observed
frequencies, which needs to be investigated. There is still some
difference between the observed and theoretical frequencies of the f-mode
at the level of $0.5\;\mu$Hz.
If these differences are real then the simplest interpretation would be
that the solar radius needs to be decreased by about 203 km as compared to
the standard value. This error is sufficiently large to affect the
standard solar models and the corresponding inversion results at a level
which is much larger than the statistical errors in inversions.
With availability of better data, f-mode frequencies
can be used to provide an independent estimate of the solar radius
and its variation with time.

The f-mode splittings have also been determined from the GONG power
spectra for $\ell\le150$. These splittings provide
an independent measure of solar surface rotation rate, which appears
to be close to that obtained from doppler measurements.

\acknow{
I am thankful to the National Solar Observatory for hospitality during
my visit to NSO where most of this work was carried out.
This work utilizes data obtained by the Global Oscillation
Network Group (GONG) project, managed by the National Solar Observatory, a
Division of the National Optical Astronomy Observatories, which is operated by
AURA, Inc. under a cooperative agreement with the National Science
Foundation. I would like to thank S. M. Chitre, F. Delmas and E. Fossat
for some useful communications.}

\begref{References}

\ref Anderson, E., Duvall, T. \& Jefferies, S. 1990, ApJ 364, 699

\ref Antia, H. M. 1996, A\&A 307, 609

\ref Bachmann, K. T., Duvall, T. L., Jr., Harvey, J. W. \& Hill, F. 1995,
ApJ 443, 837

\ref \jcd, J., D\"appen, W. et al.~1996, Science, 272, 1286.

\ref Ghosh, P., Chitre, S. M. \& Antia, H. M. 1995, ApJ 451, 851

\ref Hill, F., Stark, P. B., Stebbins, R. T. et al.~1996, Science, 272, 1292.


\ref Kosovichev, A. G. \& Schou, J. 1997, ApJ 482, L207

\ref Kosovichev, A. G. et al. 1997, Solar Phys. 171, 43.


\ref Laclare, F., Delmas, C., Coin, J. P. \& Irbah, A. 1996, Solar Phys.
166, 211

\ref Libbrecht, K. G., Woodard, M. F. \& Kaufman, J. M. 1990, ApJS 74, 1129


\ref Murawaski, K. \& Roberts, B. 1993, A\&A 272, 595


\ref Ritzwoller, M. H. \& Lavely, E. M. 1991, ApJ 369, 557

\ref Rosenthal, C. S. \& \jcd, J. 1995, MNRAS 276, 1003

\ref Rosenthal, C. S. \& Gough, D. O. 1994, ApJ 423, 488


\ref Snodgrass, H. B. 1992, in The Solar Cycle, ed. K. L. Harvey,
Astron. Soc. Pacif. Conf. Ser. Vol. 27, San Francisco, p205






\ref Wittmann, A.  1974, Solar Phys. 36, 65


\endref
\bye